\documentclass[twocolumn]{webofc}

\usepackage[varg]{txfonts}   
\usepackage{hyperref}
\usepackage{url}
\usepackage{bm}
\hypersetup{colorlinks=true,citecolor=blue,urlcolor=blue,linkcolor=blue}

\usepackage{braket}

\newcommand{\hfbtho}{{\sc hfbtho }}
\newcommand{\felix}{{\sc felix }}
\newcommand{\freya}{{\sc freya }}
\newcommand{\cgmf}{{\sc cgmf }}
\newcommand{\fifrelin}{{\sc fifrelin }}

\begin{document}
\title{Narrowing the Gap Between Theory and Evaluations: \\ 
Angular Momentum Distributions in Fission Fragments}

\author{\firstname{Petar} \lastname{Marevi\'c}\inst{1}\fnsep\thanks{\email{pmarevic@phy.hr}} \and
        \firstname{Nicolas} \lastname{Schunck}\inst{2}\fnsep \and
        \firstname{Marc} \lastname{Verriere}\inst{3,4}\fnsep
}

\institute{Department of Physics, Faculty of Science, University of Zagreb, 
HR-10000 Zagreb, Croatia 
\and
Nuclear Data and Theory Group, Nuclear and Chemical Sciences Division, 
Lawrence Livermore National Laboratory, \\ Livermore, California 94550, US
\and
CEA, DAM, DIF, 91297 Arpajon, France
\and
Universit\'e Paris-Saclay, CEA, Laboratoire Mati\`ere en Conditions Extr\^emes, 
91680 Bruy\`eres-le-Ch\^atel, France
}

\abstract{
We present a microscopic framework for predicting angular momentum distributions 
over the full range of fission fragment masses and charges.
For the neutron-induced fission of $^{235}$U and
$^{239}$Pu, the obtained distributions exhibit a pronounced sawtooth
pattern in average values, reveal a substantial isobaric dependence, 
and reproduce experimental photon multiplicities without adjustable parameters.
These results demonstrate that microscopic theory
is gradually becoming quantitatively competitive with phenomenological models.
}
\maketitle
\section{Introduction}
\label{intro}
Nuclear fission is of critical importance for both
basic science and applications \cite{talou2023}.
A longstanding goal of nuclear theory has been to describe
fission
based on our best understanding of internucleon forces
and many-body techniques, an approach known 
as microscopic modeling. 
Unfortunately, the complexity of the phenomenon
has created
a substantial gap between fundamental 
theory and applications, and predictions
of microscopic
models have long been quantitatively
inferior
to those of
phenomenological models.
However, thanks to an unprecedented increase
in computing capabilities, the last two decades
have brought rapid development of microscopic
models \cite{schunck2022}.
Today, microscopic theory provides qualitative 
insights into the fundamental 
fission mechanism and is increasingly capable of 
making reliable quantitative predictions. 

When a nucleus splits, the two fission fragments (FFs)
are typically highly excited and carry a distribution
of angular momentum.
Understanding the origin and nature of this 
rotation
presents a fundamental open question.
Moreover, the angular momentum of FFs largely drives their subsequent decay,
causing the anisotropy of neutron emission and modifying
photon multiplicities.
Recently, there has been a renaissance in theoretical studies
of the angular momentum of FFs
\cite{bulgac2021,marevic2021,bulgac2022a,scamps2022,scamps2023b,
randrup2021,stetcu2021,randrup2022a,scamps2025},
which was motivated in part by new high-resolution spectroscopy measurements at 
ALTO \cite{wilson2021}. Many aspects of the phenomenon were elucidated, 
but the studies largely remained focused on a narrow range
of FF masses around the most probable fragmentation. However, simulating the decay
of FFs in statistical reaction theory
models, such as \cgmf \cite{talou2021}, \freya \cite{verbeke2018} 
or \fifrelin \cite{litaize2015},
requires, among other ingredients, knowledge of the angular momentum
distributions for the full range of fragment masses and charges.
In the absence of a robust microscopic framework, these
simulations currently rely
on distributions extracted from simple phenomenological models, 
whose free parameters are adjusted to reproduce experimental data 
on fission spectra.

Addressing this challenge, we recently developed a microscopic framework
capable of predicting angular momentum distributions 
over the full range of FF masses and charges.
The model was applied to neutron-induced
fission of $^{239}$Pu and $^{235}$U \cite{marevic2026},
and the obtained
distributions
were made publicly available \cite{marevic2026_data}. 
In Sec.~\ref{sec:theory} we outline
the basic tenets of the model.
In Sec.~\ref{sec:results} we briefly discuss 
the results of \cite{marevic2026}, with a focus
on the isobaric dependence of the distributions
and their impact on fission spectra predictions. 
In Sec.~\ref{sec:conclusion}
we conclude and discuss future work.

\section{Theoretical model}
\label{sec:theory}

Energy density functional (EDF) theory \cite{schunck2019}
is currently the only microscopic framework
applicable to fission \cite{schunck2022}.
A typical EDF is defined by
around ten free parameters whose values
were previously adjusted to experimental data,
mostly on ground-state properties of a chosen set
of nuclei. The framework is applicable
across the entire nuclide chart and is fully agnostic
to specific properties of FFs, which
endows it with particular predictive power.

In this work, we use the \hfbtho computational framework \cite{marevic2022},
based on Skyrme EDFs and the expansion
of single-particle wave functions in the axially symmetric
harmonic oscillator basis.
First, we perform a series
of constrained Hartree-Fock-Bogoliubov (HFB)
calculations, using the
SkM* EDF and a mixed volume-surface
contact pairing force.
In this way, we generate
a set of scission configurations in the even-even
compound system, representing an odd-$N$ and even-$Z$ nucleus
that captured an incident neutron.
These configurations are characterized by different values 
of the collective variables $\bm{q} = (q_{20},q_{30},q_{N})$.
The quadrupole moment $q_{20}$ and
octupole moment $q_{30}$ quantify the
elongation and the mass asymmetry of the nuclear shape, respectively,
while the neck value $q_N \in [1, 3]$
measures the number of nucleons in a thin
neck connecting the two pre-fragments.
In addition, the configurations have 
axially symmetric density
profiles and are dumbbell shaped.
Consequently, they can be 
divided into left $(z < z_N)$ 
and right $(z > z_N)$ fission fragment, 
where $z_N$ locates the minimum
of density profile between the fragments.

For both the left $(F=l)$ and the right $(F=r)$ 
fission fragment
in each scission configuration $\ket{\Phi^S_{\bm{q}}}$, we use projection
techniques to extract
the distribution
\begin{equation}
\mathbb{P}_F(J_F, N_F, Z_F | N_0, Z_0, \bm{q}) = 
\frac{\braket{\Phi^S_{\bm{q}}
| \hat{P}^{J_F} \hat{P}^{N_F, Z_F} \hat{P}^{N_0, Z_0}|
\Phi^S_{\bm{q}}}}{\braket{\Phi^S_{\bm{q}} | P^{N_0,Z_0}  |\Phi^S_{\bm{q}}}},
\end{equation}
i.e. the probability that the fragment has an angular momentum $J_F$, 
and numbers of neutrons $N_F$ and protons $Z_F$, given
that the total system has
the correct number of neutrons $N_0$
and protons $Z_0$. Here, $\hat{P}^{J_F}$ and  $\hat{P}^{N_F, Z_F}$
are the operators projecting on
good quantum numbers in FFs,
and $P^{N_0, Z_0}$ is the operator
projecting
on good nucleon numbers
in the total system \cite{marevic2026}.

In the next step, we estimate
the probability $F(\bm{q})$ of populating
each scission configuration by employing 
the time-dependent
generator coordinate method (TDGCM)
with the Gaussian overlap approximation (GOA) \cite{verriere2020}.
Within the TDGCM+GOA framework,
the nuclear wave function is given
by a linear superposition
\begin{equation}
\ket{\Psi(t)} = \sum_q f_{\bm{q}}(t) \ket{\Phi_{\bm{q}}},
\label{eq:tdgcm}
\end{equation}
where $\ket{\Phi_{\bm{q}}}$ are HFB states on an adiabatic
$\bm{q} = (q_{20},q_{30})$ potential energy surface (PES),
and $f_\mathbf{q}(t)$ are the mixing functions that
are to be determined. Applying the time-dependent
variational principle to Eq.~\eqref{eq:tdgcm},
and assuming the GOA,
yields local, Schr{\"o}dinger-like equation
for mixing functions.
This equation can be solved using the \felix package \cite{regnier2018}.
From these solutions, one can readily calculate the
$F(\bm{q})$ probability.

Finally, combining
the two probabilities gives
\begin{equation}
\mathbb{P}_F(J_F, N_F, Z_F | N_0, Z_0) = 
\sum_{\bm{q}} F(\bm{q}) \mathbb{P}_F(J_F, N_F, Z_F | N_0, Z_0, \bm{q}).
\end{equation}
The final distribution is then given by the sum
\begin{equation}
\mathbb{P}(J_F, N_F, Z_F | N_0, Z_0)
= \sum_{F=l,r}\mathbb{P}_F(J_F, N_F, Z_F | N_0, Z_0).
\end{equation}
By fixing the values of $N_F=N_F^0$ and $Z_F=Z_F^0$
and renormalizing, we can obtain 
angular momentum distribution in the
corresponding FF,
$\mathbb{P}(J_F|N_F^0,Z_F^0,N_0,Z_0)$. 
On the other hand,
marginalization over angular momentum
$J_F$
gives a distribution in neutron
and proton numbers in FFs, i.e. the preneutron fission
yields. Furthermore, by modifying the energy of the initial 
wave packet in TDGCM+GOA, we can place a lower bound on the 
impact of different incident neutron energies $E_n$
on predictions of the model. 
To better describe the 
effect of varying $E_n$, one would need to
propagate the wave packet on top of a PES
that accounts for the excitation effects,
which is beyond the scope of this work.
Other limitations of the
model include disregarding
the $K \neq 0$ rotational components
and neglecting the
effect of intrinsic excitations of scission
configurations beyond
nuclear deformation.

\section{Neutron-induced fission of $^{235}$U and $^{239}$Pu}
\label{sec:results}

In Ref.~\cite{marevic2026}, we considered
$384$ ($404$) scission configurations in
$^{236}$U$^*$ ($^{240}$Pu$^*$), yielding
a total of $1444$ ($1601$) FFs.
Overall, at $E_n = 1$~MeV, we calculated 
angular momentum
distributions for $381$ even-even FFs 
in $^{236}$U$^*$ and $412$ in $^{240}$Pu$^*$.
These
numbers vary only slightly with $E_n$
in our model.
Generally, modifying
the initial wave packet energy only marginally impacts
model predictions. 

Several interesting features were discussed in
detail.
Notably, the model predicts a pronounced sawtooth pattern 
for the average angular momentum of FFs as a 
function of their mass,
as well as a weak correlation
between the angular momentum magnitudes
of the two partner FFs,
consistent with experiment \cite{wilson2021}.
It also reveals a clear signature of shell effects in FFs, 
predicts a strong correlation between
FF angular momentum and deformation, and
provides a reasonable description of fission yields.
In this work,
we focus on two pertinent features of
angular momentum distributions: their isobaric
dependence and impact on fission spectra.

\begin{figure}[h]
\centering
\includegraphics[width=\columnwidth,clip]{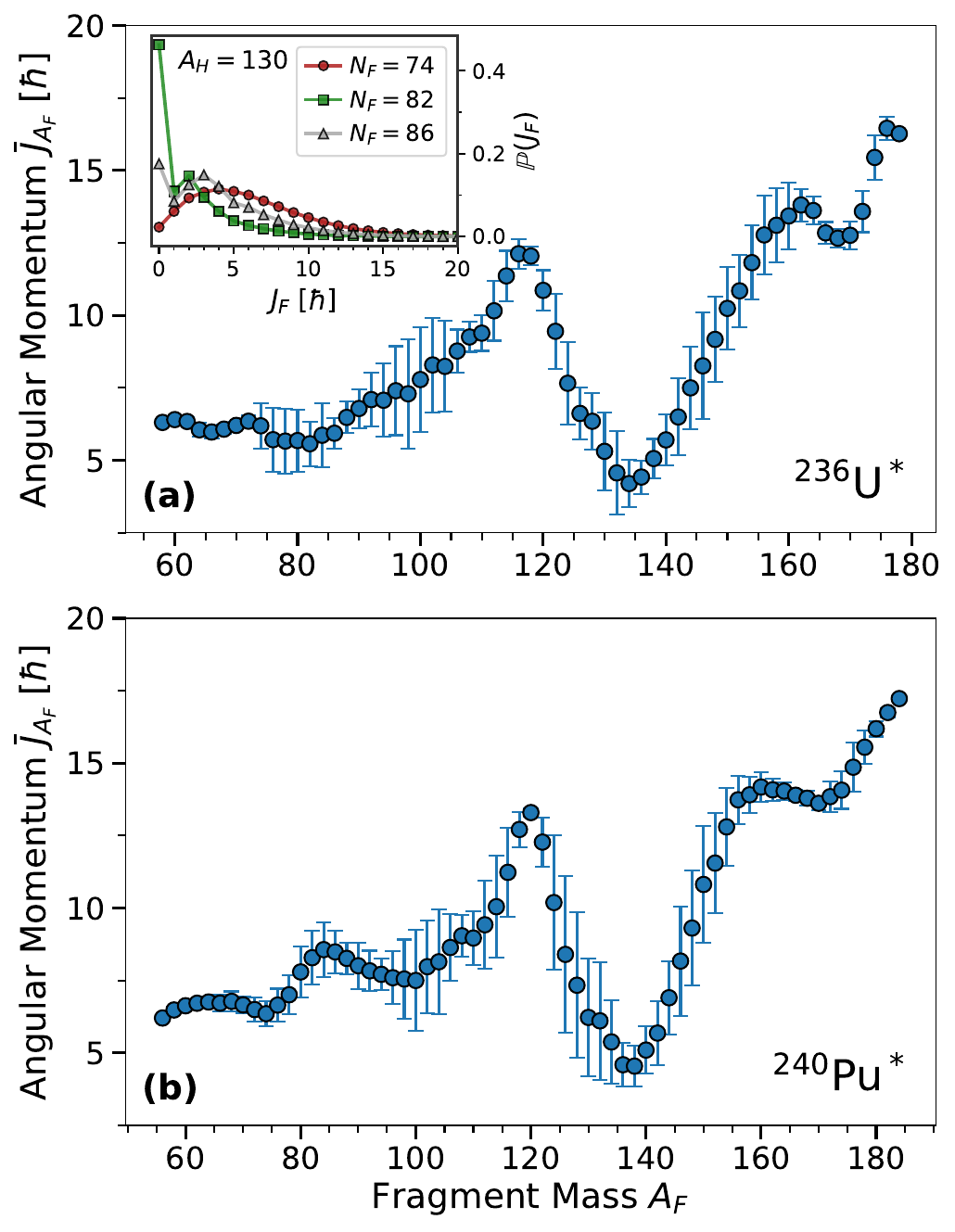}
\caption{Average angular momentum of FFs in isobaric 
chains as a function of the FF
mass $A_F$, for $^{236}$U$^*$ (a) and $^{240}$Pu$^*$ (b).
The error bars show $1\sigma$
spread within each chain.
The inset shows distributions in three $A_H = 130$ isobars 
for fission of $^{236}$U$^*$.}
\label{fig-1}       
\end{figure}

To quantify the isobaric dependence of distributions,
we calculate the average
angular momentum in each isobaric chain,
$
\bar{J}_{A_F} = \sum_{N_F + Z_F = A_F} \bar{J}_{N_F, Z_F}$,
where the sum runs over all even-even FFs that satisfy
$N_F + Z_F = A_F$, and $\bar{J}_{N_F, Z_F}$
is an average angular momentum of each FF.
It can be estimated
as $\bar{J}_{N_F,Z_F}(\bar{J}_{N_F,Z_F}+1) = \sum_{J_F} J_F (J_F+1) \mathbb{P}(J_F)$,
where $\mathbb{P}(J_F)$ is short
for $\mathbb{P}(J_F|N_F^0,Z_F^0,N_0,Z_0)$.
In Fig.~\ref{fig-1} we show $\bar{J}_{A_F}$
as a function of $A_F$ for both nuclei,
as well as the $1\sigma$ spread within each isobaric chain.
In addition to exhibiting a pronounced sawtooth pattern,
the $\bar{J}_{A_F}$ shows a clear isobaric dependence. 
The effect is particularly
strong in both FFs near the most probable fragmentation
($A_H \approx 136-140$), where the $1\sigma$ spread
can be as large as $\pm 3\hbar$. 
The isobaric dependence
is further illustrated in the inset
of the upper panel, where we show distributions
in three FFs
with $A_H = 130$.
While the magic $N_F = 82$ FF has a pronounced maximum at 
$J_F = 0$, moving away from the closed shell in either
direction significantly increases
the angular momentum of FFs.
This effect, however, is not limited to
FFs near shell closures. Even the distributions
of well-deformed FFs can exhibit large
variations within isobaric chains,
such as those in the $A_H = 150$ chain \cite{marevic2026}.

On the other hand, phenomenological models typically sample
the FF angular momentum from a distribution of the form
\begin{equation}
p(J_F) \propto (2J_F+1) \exp\Big(-\frac{1}{2}\frac{J_F (J_F+1)}{B^2(Z_F, A_F, T_F)}\Big),
\label{eq:phenomeno}
\end{equation}
where $B^2(Z_F, A_F, T_F)$ encodes a possible dependence on FF charge, 
mass, and temperature, and includes adjustable parameters that are 
fixed by fitting the model predictions for fission spectra to 
experimental data. However, with the exception of 
\cgmf \cite{talou2021}, such models typically consider only 
variations with $A_F$ and $T_F$, while any isobaric dependence 
is entirely disregarded. This choice is rather successful
for a small number of nuclei where ample experimental
data are available, but may fall short in cases
where such data are scarce.
Our result thus illustrates how microscopic theory 
could be useful in informing phenomenological models
and potentially improving their predictive power.

Phenomenological models such as the one of Eq.~\eqref{eq:phenomeno} 
are typically used to set the initial angular momentum distributions 
of primary FFs for simulating their decay within statistical reaction theory.
The adjustable parameters of Eq.~\eqref{eq:phenomeno} are carefully adjusted to
reproduce experimental fission spectra.
To test the quality of our microscopic distributions
and assess their impact on fission spectra,
we used them as inputs to the state-of-the-art
\cgmf code \cite{talou2021}, while keeping all the other 
initial conditions unchanged at their \cgmf default value.

\begin{figure}[h]
\centering
\includegraphics[width=\columnwidth,clip]{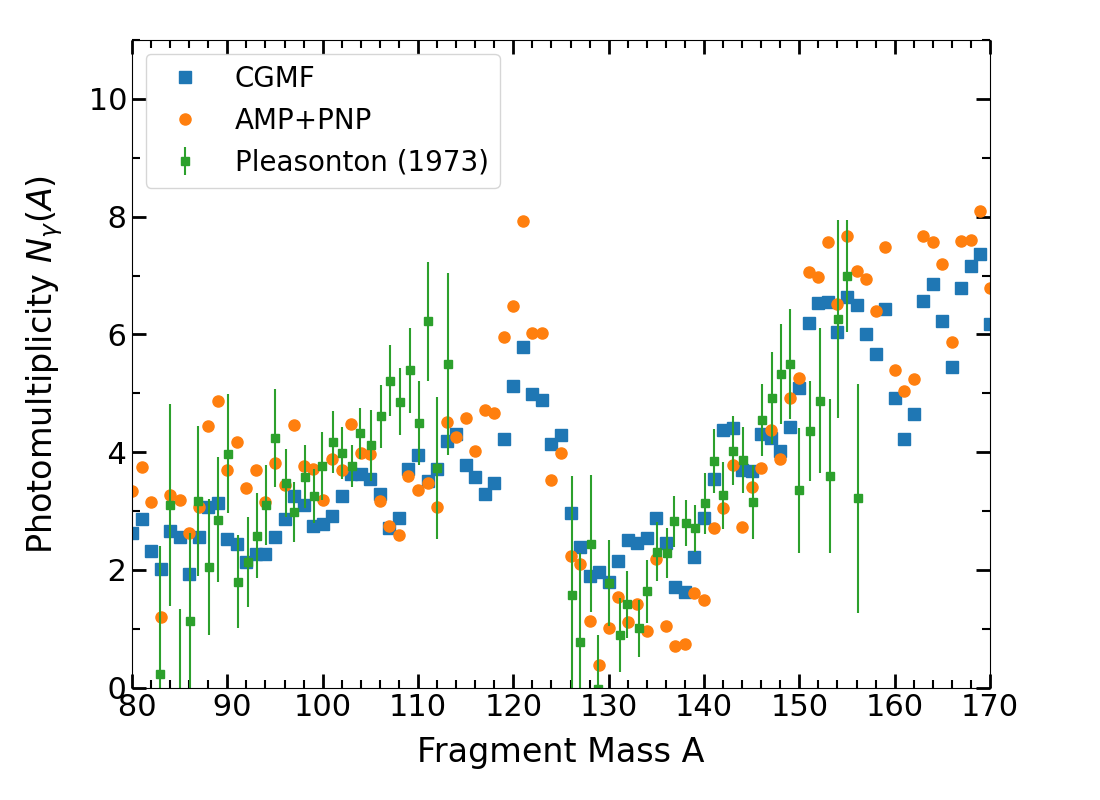}
\caption{Photon multiplicities from FFs in fission
of $^{240}$Pu$^*$, calculated 
with the \cgmf code. FFs are deexcited using either microscopic
or default phenomenological angular momentum
distributions, while keeping
all other initial conditions equal.
Experimental data are taken from Ref.~\cite{pleasanton1973}.}
\label{fig-2}  
\end{figure}

In Fig.~\ref{fig-2}, we compare the 
\cgmf
photon multiplicities obtained with microscopic
distributions
to those obtained using the default phenomenological
distributions. It is remarkable that
the experimental multiplicities are essentially
reproduced with microscopic
distributions without
adjustable
parameters. The total photon multiplicity
drops by about $0.2$ photons, from $6.28$ to $6.08$, 
a difference which is within a
typical experimental
margin of error. We also observe
that neutron
multiplicities are only slightly impacted,
dropping from $2.86$ to $2.80$.
This is, to the best of our knowledge,
the most prominent example thus far of microscopic
theory being quantitatively competitive
with phenomenological models.

\section{Conclusion and outlook}
\label{sec:conclusion}

Microscopic fission theory has matured to the point
of providing reliable quantitative predictions.
Notably, angular momentum distributions in FFs
can now be predicted for the full range of
masses and charges,
offering valuable guidance for phenomenological models
and pertinent inputs for FF decay models.
Future work will likely pursue two complementary
avenues. On the one hand, modeling
of the angular momentum can be further refined, for
example by including the intrinsic excitation
effects for the full range
of FFs and/or restoring other symmetries, such
as parity. On the other hand, dedicated efforts
are being made to improve microscopic predictions
of other essential ingredients to statistical
reaction theory models,
such as fission yields and FF excitation energies,
whose quality is still not at the level
of phenomenological models.
Eventual success of these efforts may
pave the way to the modeling of
FF decay based on robust inputs from microscopic theory.

\section*{Acknowledgements}

The work of P.M. was funded by the 
European Union's Horizon Europe research and innovation program 
under the Marie Skłodowska-Curie Actions Grant Agreement No. 101149053. 
Support for this work was partly provided through Scientific Discovery 
through Advanced Computing (SciDAC) program funded by 
U.S. Department of Energy, Office of Science, Advanced Scientific 
Computing Research and Nuclear Physics. This work was partly 
performed under the auspices of the US Department of Energy by the 
Lawrence Livermore National Laboratory under Contract No. DE-AC52-07NA27344. 
Computing support came from the Lawrence Livermore National Laboratory 
Institutional Computing Grand Challenge program.

\bibliography{bibliography}

@article{schunck2022,
title = {Theory of nuclear fission},
journal = {Progress in Particle and Nuclear Physics},
volume = {125},
pages = {103963},
year = {2022},
issn = {0146-6410},
doi = {https://doi.org/10.1016/j.ppnp.2022.103963},
url = {https://www.sciencedirect.com/science/article/pii/S0146641022000242},
author = {Nicolas Schunck and David Regnier}
}

@article{bulgac2021,
  title = {Fission Fragment Intrinsic Spins and Their Correlations},
  author = {Bulgac, Aurel and Abdurrahman, Ibrahim and Jin, Shi and Godbey, Kyle and Schunck, Nicolas and Stetcu, Ionel},
  journal = {Phys. Rev. Lett.},
  volume = {126},
  issue = {14},
  pages = {142502},
  numpages = {7},
  year = {2021},
  month = {Apr},
  publisher = {American Physical Society},
  doi = {10.1103/PhysRevLett.126.142502},
  url = {https://link.aps.org/doi/10.1103/PhysRevLett.126.142502}
}

@article{marevic2026,
  title = {Microscopic theory of angular momentum distributions across the full range of fission fragments},
  author = {Marevi\ifmmode \acute{c}\else \'{c}\fi{}, Petar and Schunck, Nicolas and Verriere, Marc},
  journal = {Phys. Rev. C},
  volume = {113},
  issue = {1},
  pages = {014612},
  numpages = {18},
  year = {2026},
  month = {Jan},
  publisher = {American Physical Society},
  doi = {10.1103/yr2c-nvf3},
  url = {https://link.aps.org/doi/10.1103/yr2c-nvf3}
}

@misc{scamps2025,
      title={Uncertainty Principle and Angular Momentum Generation in Microscopic Fission Models}, 
      author={G. Scamps and A. Guilleux and D. Regnier and A. Bernard},
      year={2025},
      eprint={2512.02207},
      archivePrefix={arXiv},
      primaryClass={nucl-th},
      url={https://arxiv.org/abs/2512.02207}, 
}

@article{marevic2021,
  title = {Angular momentum of fission fragments from microscopic theory},
  author = {Marevi\ifmmode \acute{c}\else \'{c}\fi{}, Petar and Schunck, Nicolas and Randrup, J\o{}rgen and Vogt, Ramona},
  journal = {Phys. Rev. C},
  volume = {104},
  issue = {2},
  pages = {L021601},
  numpages = {6},
  year = {2021},
  month = {Aug},
  publisher = {American Physical Society},
  doi = {10.1103/PhysRevC.104.L021601},
  url = {https://link.aps.org/doi/10.1103/PhysRevC.104.L021601}
}

@dataset{marevic2026_data,
  author       = {Marević, Petar and Schunck, Nicolas},
  title        = {Microscopic Angular Momentum Distributions in Fragments
                  for Neutron-Induced Fission of {U}-235 and {Pu}-239},
  year         = {2025},
  publisher    = {Zenodo},
  version      = {v1},
  doi          = {10.5281/zenodo.17303186},
  url          = {https://doi.org/10.5281/zenodo.17303186},
  license      = {CC-BY-4.0}
}

@article{wilson2021,
  title = {{Angular momentum generation in nuclear fission}},
  author = {Wilson et al., J. N.},
  journal = {Nature},
  volume = {590},
  pages = {566-570},
  year = {2021},
  doi = {https://doi.org/10.1038/s41586-021-03304-w},
  url = {10.1038/s41586-021-03304-w}
}

@article{scamps2023b,
  title = {Spatial orientation of the fission fragment intrinsic spins and their correlations},
  author = {Scamps, Guillaume and Abdurrahman, Ibrahim and Kafker, Matthew and Bulgac, Aurel and Stetcu, Ionel},
  journal = {Phys. Rev. C},
  volume = {108},
  issue = {6},
  pages = {L061602},
  numpages = {6},
  year = {2023},
  month = {Dec},
  publisher = {American Physical Society},
  doi = {10.1103/PhysRevC.108.L061602},
  url = {https://link.aps.org/doi/10.1103/PhysRevC.108.L061602}
}

@book{talou2023,
  title     = {{Nuclear Fission: Theories, Experiments and Applications}},
  publisher = {Springer},
  year      = {2023},
  editor    = {Talou, P. and Vogt, R.}
}

@article{bulgac2022a,
  title = {Fragment Intrinsic Spins and Fragments' Relative Orbital Angular Momentum in Nuclear Fission},
  author = {Bulgac, Aurel and Abdurrahman, Ibrahim and Godbey, Kyle and Stetcu, Ionel},
  journal = {Phys. Rev. Lett.},
  volume = {128},
  issue = {2},
  pages = {022501},
  numpages = {6},
  year = {2022},
  month = {Jan},
  publisher = {American Physical Society},
  doi = {10.1103/PhysRevLett.128.022501},
  url = {https://link.aps.org/doi/10.1103/PhysRevLett.128.022501}
}

@article{randrup2021,
  title = {Generation of Fragment Angular Momentum in Fission},
  author = {Randrup, J\o{}rgen and Vogt, Ramona},
  journal = {Phys. Rev. Lett.},
  volume = {127},
  issue = {6},
  pages = {062502},
  numpages = {5},
  year = {2021},
  month = {Aug},
  publisher = {American Physical Society},
  doi = {10.1103/PhysRevLett.127.062502},
  url = {https://link.aps.org/doi/10.1103/PhysRevLett.127.062502}
}

@article{randrup2022a,
  title = {Probing fission fragment angular momenta by photon measurements},
  author = {Randrup, J\o{}rgen and D\o{}ssing, Thomas and Vogt, Ramona},
  journal = {Phys. Rev. C},
  volume = {106},
  issue = {1},
  pages = {014609},
  numpages = {10},
  year = {2022},
  month = {Jul},
  publisher = {American Physical Society},
  doi = {10.1103/PhysRevC.106.014609},
  url = {https://link.aps.org/doi/10.1103/PhysRevC.106.014609}
}

@article{scamps2022,
  title = {Microscopic description of the torque acting on fission fragments},
  author = {Scamps, Guillaume},
  journal = {Phys. Rev. C},
  volume = {106},
  issue = {5},
  pages = {054614},
  numpages = {8},
  year = {2022},
  month = {Nov},
  publisher = {American Physical Society},
  doi = {10.1103/PhysRevC.106.054614},
  url = {https://link.aps.org/doi/10.1103/PhysRevC.106.054614}
}

@article{stetcu2021,
  title = {{Angular Momentum Removal by Neutron and $\ensuremath{\gamma}$-Ray Emissions during Fission Fragment Decays}},
  author = {Stetcu, I. and Lovell, A. E. and Talou, P. and Kawano, T. and Marin, S. and Pozzi, S. A. and Bulgac, A.},
  journal = {Phys. Rev. Lett.},
  volume = {127},
  issue = {22},
  pages = {222502},
  numpages = {6},
  year = {2021},
  month = {Nov},
  publisher = {American Physical Society},
  doi = {10.1103/PhysRevLett.127.222502},
  url = {https://link.aps.org/doi/10.1103/PhysRevLett.127.222502}
}

@article{verbeke2018,
title = "{Fission Reaction Event Yield Algorithm FREYA 2.0.2}",
journal = "Computer Physics Communications",
volume = "222",
pages = "263 - 266",
year = "2018",
issn = "0010-4655",
doi = "https://doi.org/10.1016/j.cpc.2017.09.006",
url = "http://www.sciencedirect.com/science/article/pii/S001046551730293X",
author = "J. M. Verbeke and J. Randrup and R. Vogt"
}

@article{talou2021,
title = {Fission fragment decay simulations with the {CGMF} code},
journal = {Computer Physics Communications},
volume = {269},
pages = {108087},
year = {2021},
issn = {0010-4655},
doi = {https://doi.org/10.1016/j.cpc.2021.108087},
url = {https://www.sciencedirect.com/science/article/pii/S0010465521001995},
author = {P. Talou and I. Stetcu and P. Jaffke and M.E. Rising and A.E. Lovell and T. Kawano}
}

@article{litaize2015,
  title = {Fission modelling with {FIFRELIN}},
  author = {Litaize, O. and Serot, O. and Berge, L.},
  journal = {Eur. Phys. J. A},
  volume = {51},
  pages = {177},
  year = {2015},
  doi = {10.1140/epja/i2015-15177-9},
  url = {https://link.springer.com/article/10.1140/epja/i2015-15177-9#citeas}
}

@Book{schunck2019,
  editor    = {Schunck, N.},
  publisher = {{IOP Publishing}},
  title     = {{Energy Density Functional Methods for Atomic Nuclei.}},
  year      = {2019},
  address   = {{Bristol, UK}},
  series    = {{{IOP Expanding Physics}}},
  doi       = {10.1088/2053-2563/aae0ed}
}

@article{marevic2022,
title = {{Axially-deformed solution of the Skyrme-Hartree-Fock-Bogoliubov equations using the transformed harmonic oscillator basis (IV) hfbtho (v4.0): A new version of the program}},
journal = {Computer Physics Communications},
volume = {276},
pages = {108367},
year = {2022},
issn = {0010-4655},
doi = {https://doi.org/10.1016/j.cpc.2022.108367},
url = {https://www.sciencedirect.com/science/article/pii/S0010465522000868},
author = {P. Marević and N. Schunck and E.M. Ney and R. {Navarro Pérez} and M. Verriere and J. O'Neal}
}

@article{regnier2018,
title = {{FELIX-2.0: New version of the finite element solver for the time dependent generator coordinate method with the Gaussian overlap approximation}},
journal = {Computer Physics Communications},
volume = {225},
pages = {180-191},
year = {2018},
issn = {0010-4655},
doi = {https://doi.org/10.1016/j.cpc.2017.12.007},
url = {https://www.sciencedirect.com/science/article/pii/S0010465517304125},
author = {D. Regnier and N. Dubray and M. Verrière and N. Schunck}
}

@article{pleasanton1973,
title = {{Prompt $\gamma$-rays emitted in the thermal-neutron induced fission of $^{233}$U and $^{239}$Pu}},
journal = {Nuclear Physics A},
volume = {213},
number = {2},
pages = {413-425},
year = {1973},
issn = {0375-9474},
doi = {https://doi.org/10.1016/0375-9474(73)90161-9},
url = {https://www.sciencedirect.com/science/article/pii/0375947473901619},
author = {Frances Pleasonton}
}

@article{verriere2020,
  title = {The {{Time-Dependent Generator Coordinate Method}} in {{Nuclear Physics}}},
  author = {Verriere, Marc and Regnier, David},
  year = {2020},
  journal = {Front. Phys.},
  volume = {8},
  pages = {1},
  publisher = {Frontiers},
  doi = {10.3389/fphy.2020.00233},
  langid = {english}
}

\end{document}